\documentclass[useAMS,usenatbib]{mn2e}
\usepackage[dvips]{epsfig}
\usepackage[]{natbib}
\title[M81]{Gas Streaming Motions towards the Nucleus of M\,81}
\author[A. Schnorr M\"uller et al.]
  {Allan Schnorr M\"uller,$^1$ Thaisa Storchi-Bergmann,$^1$ Rogemar A. Riffel,$^2$ 
\newauthor Fabricio Ferrari,$^3$ J. E. Steiner,$^4$ David J. Axon,$^{5,6}$ Andrew Robinson$^5$\\
  $^1$Instituto de F\'isica, Universidade Federal do Rio Grande do Sul, 91501-970, Porto Alegre, RS, Brazil\\
  $^2$Universidade Federal de Santa Maria, Departamento de F\'isica, 97105-900, Santa Maria, RS, Brazil\\
  $^3$Universidade Federal do Pampa, Campus Bag\'e, 96412-420, Bag\'e, RS, Brazil\\
  $^4$Instituto de Astronomia, Geof\'\i{}sica e Ci\^encias Atmosf\'ericas, Universidade de S\~ao Paulo, 05508-900, S\~ao Paulo, SP, Brazil\\
  $^5$Physics Department, Rochester Institute of Technology, Rochester, New York 14623, USA\\
  $^6$University of Sussex, School of Mathematical and Physical Sciences, University of Sussex, Sussex House, Brighton, BN1 9RH, UK\\}
\date{Released 2010}

\pagerange{\pageref{firstpage}--\pageref{lastpage}} \pubyear{2010}

\begin{document}

\label{firstpage}

\maketitle

\begin{abstract}
We present two-dimensional stellar and gaseous kinematics of the inner 120\,$\times$\,250\,pc$^2$ of the Liner/Seyfert\,1 galaxy M\,81, from optical spectra obtained with the GMOS integral field spectrograph on the Gemini North telescope at a spatial resolution of $\approx$\,10\,pc. The  stellar velocity field shows circular rotation and overall is very similar to the published large scale velocity field, but deviations are observed close to the minor axis which can be attributed to stellar motions possibly associated to a nuclear bar. The stellar velocity dispersion of the bulge is 162\,$\pm$\,15\,km\,s$^{-1}$, in good agreement with previous measurements and leading to a black hole mass of M$_{BH}$\,=\,5.5$_{-2.0}^{+3.6}$\,$\times$10$^{7}$\,M$_{\odot}$ based on the M$_{BH}$-$\sigma$ relationship. The gas kinematics is  dominated by non-circular motions and the subtraction of the stellar velocity field reveals blueshifts of $\approx$\,$-100$\,km\,s$^{-1}$ on the far side of the galaxy and a few redshifts on the near side.  These characteristics can be interpreted in terms of streaming towards the center if the gas is in the plane. On the basis of the observed velocities and geometry of the flow, we estimate a mass inflow rate in ionized gas of $\approx$\,4.0\,$\times$\,10$^{-3}$\,M$_{\odot}$\,year$^{-1}$, which is of the order of the accretion rate necessary to power the LINER nucleus of M\,81. We have also applied the technique of Principal Component Analysis (PCA) to our data, which reveals the presence of a rotating nuclear gas disk within $\approx$\,50\,pc from the nucleus and a compact outflow, approximately perpendicular to the disk. The PCA combined with the observed gas velocity field shows that the nuclear disk is being fed by gas circulating in the galaxy plane. The presence of the outflow is supported by a compact jet seen in radio observations at a similar orientation, as well as by an enhancement of the [OI]/H$\alpha$ line ratio, probably resulting from shock excitation of the circumnuclear gas by the radio jet. With these observations we are thus resolving both the feeding -- via the nuclear disk and observed gas inflow, and the feedback -- via the outflow, around the low-luminosity active nucleus of M\,81.

\end{abstract}


\section{Introduction}

It is widely accepted that the radiation emitted by an active galactic nucleus (AGN) is a result of accretion onto the central supermassive black hole (hereafter SMBH). However, the exact nature of the mechanisms responsible for transfering mass from galactic scales down to nuclear scales to feed  the SMBH is still an open question. Theoretical studies and simulations \citep{shlosman90,emsellem03,knapen05,emsellem06} have shown that non-axisymmetric potentials efficiently promote gas inflow towards the inner regions \citep{englmaier04}. Imaging studies have revealed that structures such as small-scale disks or nuclear bars and associated spiral arms are frequently observed in the inner kiloparsec of active galaxies \citep{erwin99,pogge02,laine03}, although \citet{shaw93} raises important points about the identification of nuclear bars.

While bars can be effective in transporting gas into the inner few hundred parsecs, the fundamental problem of how gas gets from there down to the SMBH has remained unsolved. Recently \citet{lopes07} have shown a strong correlation between the presence of nuclear dust structures (filaments, spirals and disks) and activity in galaxies. \citet{garcia-burillo05} have argued that stellar gravity torques are also a mechanism that can drive gas inwards at least to a few hundred parsecs of the nuclei. 

Inward gas streaming  on galactic scales has only been observed in  a few objects. \citet{adler96} found gas inflows along the spirals arms of  M\,81 and \citet{mundell99} found inflows along the bar of NGC\,4151. An important recent development has been the recognition of streaming motions on nuclear scales in both the ionized and molecular gas. In the ionized gas, inflows have been  observed in the central region of NGC\,1097 \citep{fathi06} and NGC\,6951 \citep{thaisa07} and in molecular hydrogen (H$_2$ ) in NGC\,1068 \citep{mueller08}, NGC\,4051 \citep{riffel08}, NGC\,1097 \citep{davies09} and Mrk\,1066 \citep{rogemar11}.

With the goal of looking for more cases of inward streaming motions we began a project to map the gaseous kinematics around nearby AGN. In the present work, we present results obtained from integral field spectroscopic observations of the nuclear region of M\,81, a spiral galaxy with Hubble type SA(s)ab. At a distance of 3.5\,Mpc \citep{paturel02}, corresponding to a scale of 17\,pc\,arcsec$^{-1}$, M\,81 harbors the nearest directly observable low-luminosity AGN, classified as LINER\,/\,Seyfert 1 \citep{heckman80,peimbert81}. Because of M\,81 proximity, the AGN has been the subject of many studies. It was found to vary with time at optical \citep{bower96}, radio \citep{ho99} and X-ray wavelengths \citep{Iyomoto01}. The nuclear spectrum presents broad optical and UV emission lines \citep{peimbert81} and a featureless UV continuum \citep{ho96}. \citet{devereux97} detected a compact optical\,/\,UV continuum source, and \citet{bietenholz00} detected an ultra compact radio source associated with a radio jet. Its accretion physics was studied by \citet{markoff08}. 

The present paper is organized as follows. In Section \ref{obs} we describe the observations and
reductions. In Section \ref{results} we present the results. In section \ref{kinematics} we present the stellar and gaseous kinematics. In section \ref{fluxgas} we present the gas flux distributions and ratios, in section \ref{discuss} we discuss our results and an alternative method for the analysis of the gas kinematics using the technique of Principal Component Analysis. In Section \ref{inflow} we present an estimate of the mass inflow rate and in Section \ref{conc} we present our conclusions.

\begin{figure*}
\includegraphics[scale=0.8]{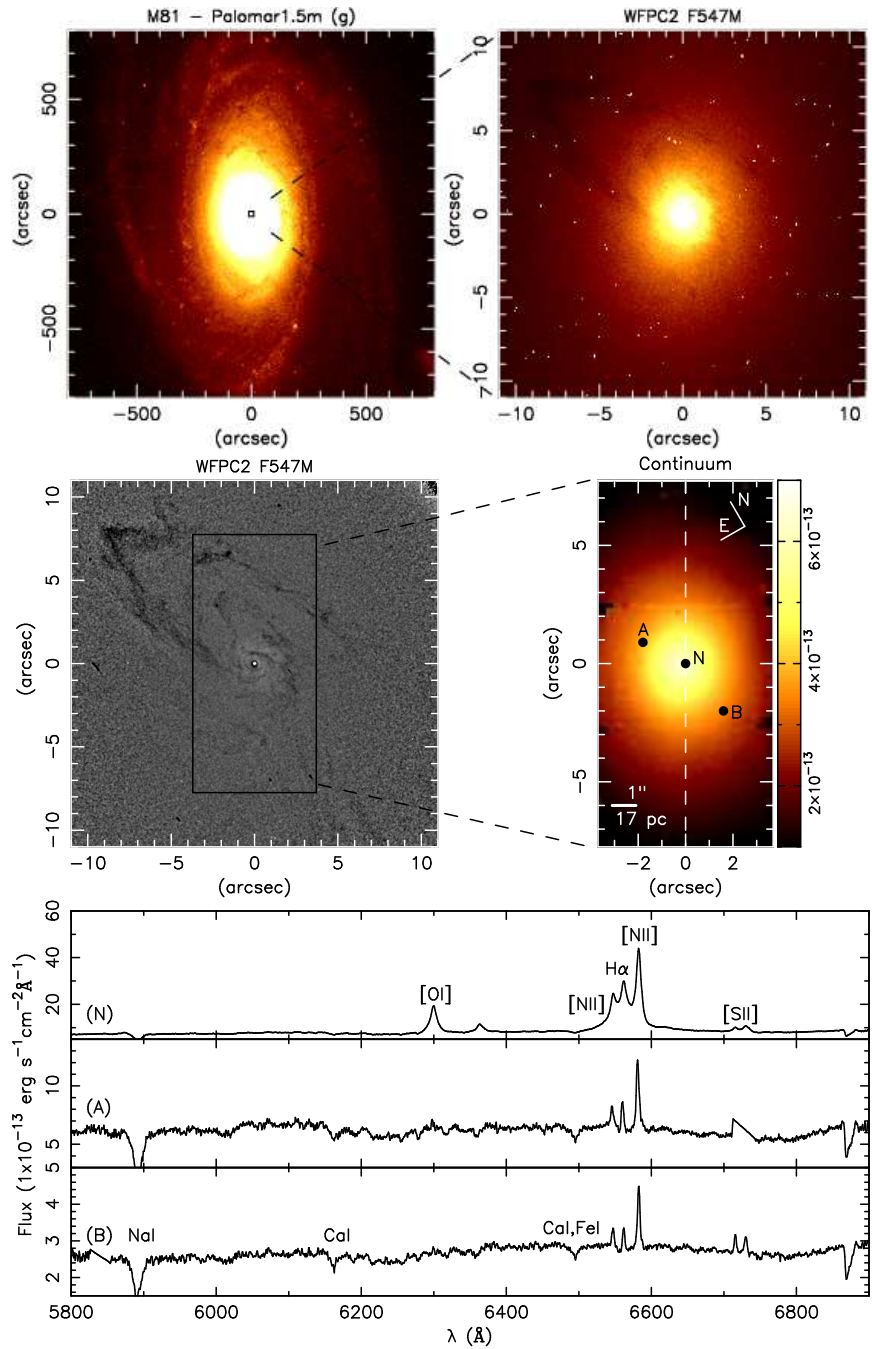}
\includegraphics[scale=0.8]{fig1b.eps}
\caption[Large scale image of M81]{Top left: large scale image of M81. Top right: WFPC2 image. Middle left: structure map. The rectangle shows the field of the IFU observation. Middle right: continuum image from the IFU spectra (flux in erg\,cm$^2$\,s$^{-1}$ per pixel). The dashed white line indicates the position of the line of nodes. Bottom: spectra corresponding to the regions marked as N, A and B in the IFU image.}
\label{fig1}
\end{figure*}

\section {Observations and Reductions}\label{obs}
The observations were obtained with the Integral Field Unit of the Gemini Multi Object Spectrograh (GMOS-IFU) at the Gemini North telescope on the night of December 31, 2006 (Gemini project GN-2006B-Q-94). The observations consisted of three adjacent IFU fields (covering 7\,$\times$\,5\,arcsec$^{2}$ each) resulting in a total angular coverage of 7\,$\times$\,15\,arcsec$^{2}$ around the nucleus, with the longest extent of the field chosen to be along the galaxy major axis as reported by \citet{goad76} (position angle PA\,=\,150\ensuremath{^\circ}). Three exposures of 191 seconds were obtained for each field, slightly shifted in order to correct for detector defects after combination of the frames. The seeing during the observation was 0.6\,\arcsec, as measured from the FWHM of a spatial profile of the galaxy extracted along the broad H$\alpha$ profile under the assumption that the broad-line region is unresolved. This corresponds to a spatial resolution at the galaxy of 10.2\,pc.

The selected wavelength range was 5600-7000\,\r{A}, in order to cover the H$\alpha$+[N\,II]\,$\lambda\lambda$6548,6583 and [S\,II]\,$\lambda\lambda$6716,6731 emission lines, observed with the grating GMOS R400-G5305 (set to central wavelength $\lambda$\,6300\,\r{A}) at a spectral resolution of R\,$\approx$\,2000.

The data reduction was performed using specific tasks developed for GMOS data in the
\textit{gemini.gmos} package as well as generic tasks in \textit{IRAF}\footnote{\textit{IRAF} is distributed by the National Optical Astronomy Observatories, which are operated by the Association of Universities for Research in Astronomy, Inc., under cooperative agreement with the National Science Foundation.}. The reduction process comprised bias subtraction, flat-fielding, trimming, wavelength calibration, sky subtraction, relative flux calibration, building of the data cubes at a sampling of 0.1\arcsec$\,\times\,$0.1\arcsec, and finally the alignment and combination of the 9 data cubes. Unfortunately, in many of the spectra the position of the [S\,II]\,$\lambda\lambda$6716,6731 emission lines on the detector coincided with the position of one of the gaps between the CCDs, thus these lines could not be measured at several locations, as is the case of the spectrum extracted from the position marked as A on Fig.\,\ref{fig1}. As our data cubes are somewhat oversampled (relative to the seeing), a Gaussian filter with FWHM\,=\,3\,pixels was applied to the centroid velocity, velocity dispersion, flux and line ratio maps in order to reduce the noise in the figures.

\section{Results}\label{results}

In Fig.\,\ref{fig1} we show in the upper left panel a large scale image of M\,81 in the $R$ band
\citep{cheng97}. In the upper right panel we show an image of the inner 22\arcsec\,$\times{}\,$22\arcsec\,\,of the galaxy obtained with the WFPC2 (Wide Field Planetary Camera 2) through the filter F547M aboard the HST. In the middle left panel we present a structure map of the WFPC2 HST image of M\,81 (see \citet{lopes07}), where a large nuclear spiral structure, as delineated by the contrast provided by the dust lanes, can be seen. The rectangle shows the field-of-view (hereafter FOV) covered by the IFU observations. In the middle right panel we present an image from our IFU observations obtained by integrating the flux within a spectral window containing only continuum. In the lower panel we present three spectra of the galaxy corresponding to locations marked as A, B and N in the IFU image and extracted within apertures of 0.1\arcsec\,$\times{}\,$0.1\arcsec.

The spectrum corresponding to the nucleus (marked as N in Fig.\,\ref{fig1}) is similar to that obtained previously with the HST \citep{bower96,devereux03}, showing [O\,I]\,$\lambda$$\lambda$\,6300,6363, [N\,II]\,$\lambda$$\lambda$6548,6583 and [S\,II]\,$\lambda$$\lambda$6717,6731 emission lines, which are broader than in the extranuclear spectra. The broad double peaked H$\alpha$ profile reported by \citet{bower96} is not  easily seen in our observations, suggesting that its flux has decreased since 1995, when that spectrum was obtained. Spectra from locations A and B show fainter emission in [N\,II], H$\alpha$, and [S\,II] on top of a strong stellar continuum.

\section{Kinematics}\label{kinematics}
\subsection{Stellar Kinematics}

\begin{figure}
\center
\includegraphics[scale=0.8]{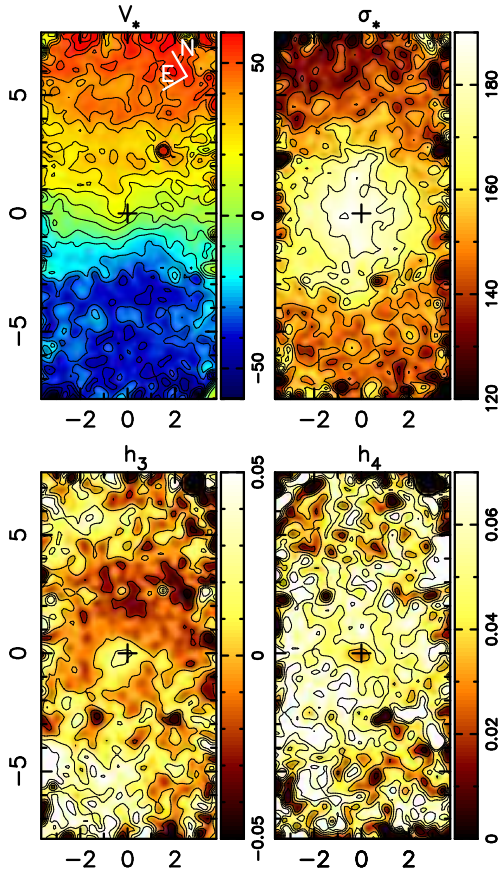}
\caption[Stellar Kinematics]{Top left: stellar centroid velocity field (km\,s$^{-1}$). Top right: stellar velocity dispersion (km\,s$^{-1}$). Bottom left: Gauss-Hermitte moment \textit{h$_{3}$}. Bottom right: Gauss-Hermitte moment \textit{h$_{4}$}.}
\label{fig2}
\end{figure}
\begin{figure}
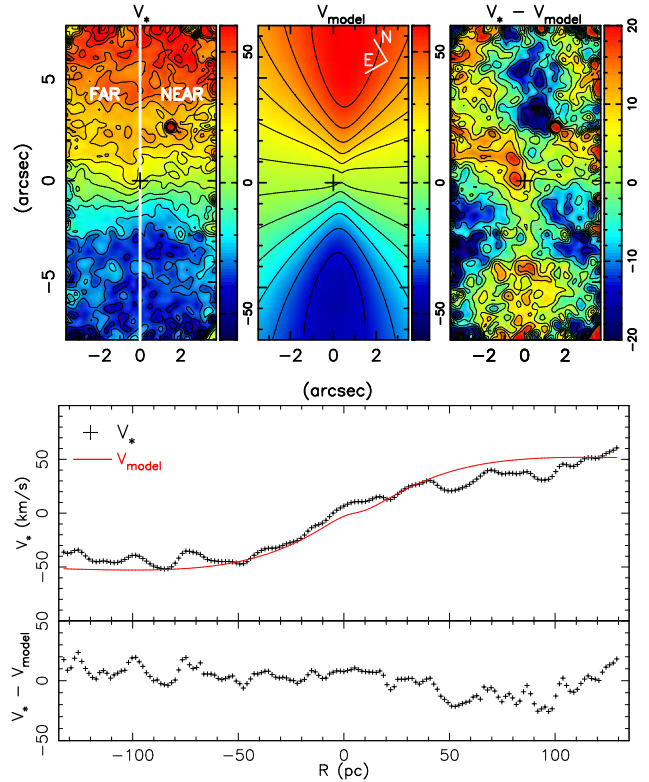

\includegraphics[scale=0.61]{fig3a.eps}
\includegraphics[scale=0.44]{fig3b.eps}
\caption[Stellar Kinematics]{At the top, from left to right: stellar centroid velocity field (km\,s$^{-1}$), model
velocity field (km\,s$^{-1}$), residuals between stellar and model velocity field (km\,s$^{-1}$). Note the change of scale between the velocity fields and the residuals The straight white line indicates the position of the line of nodes. At the bottom we show the stellar and
model velocity curve along the major axis (top) and the residuals
along the major axis (bottom)}
\label{fig3}
\end{figure}

The spectra of M\,81 have a high signal-to-noise ratio in the continuum (see Fig.\,\ref{fig1}) which has allowed us to use the stellar absortion features between 5600\,\r{A} and  7000\,\r{A} to derive the stellar kinematics. The Penalized Pixel Fitting technique (pPXF) \citep{cappellari04} was employed to obtain the stellar centroid velocity, velocity dispersion and the Gauss-Hermite moments \textit{{h$_{3}$}} and \textit{h$_{4}$} from each spectrum, using as templates the \citet{bruzual03} models. Although these models have an instrumental width ($\sigma$) of 61\,km\,s$^{-1}$ at 6300\AA, very similar to the value of 66\,km\,s$^{-1}$ of our observations, we have corrected our measurements by this small effect, which results in a decrease of $\approx$\,1.5\,km\,s$^{-1}$ in the stellar velocity dispersions. Monte Carlo simulations based on the best-fitting absorption spectra were carried out to estimate the errors in the kinematic parameters. 

The stellar centroid velocity (V$_{*}$) field (top left panel of Fig.\,\ref{fig2}) displays a rotation pattern in which the SE side of the galaxy is aproaching and the NW side is receeding. Under the assumption that the spiral arms are trailing, it can be concluded that the near side of the galaxy is the SW, and the far side is the NE. The calculated errors in the velocity measurement are $\approx\,$10\,km\,s$^{-1}$ over the whole FOV. 
The stellar velocity dispersion (top right panel of Fig.\,\ref{fig2}) decreases from decreases from $\ge\,$190\,km\,s$^{-1}$ at the nucleus to $\approx$140\,km\,s$^{-1}$ at a radial distance of $\approx\,$5\arcsec ($\approx$\,85\,pc from the nucleus), with an average value of 153\,km\,s$^{-1}$. Calculated errors are $\approx\,$15\,km\,s$^{-1}$.

The Gauss-Hermite moments $h_3$ and $h_4$ (bottom panels of Fig.\,\ref{fig2}) are measures of the skewness and the kurtosis, respectively, of the  line-of-sight velocity distribution (LOSVD). Errors are of the order of 0.03 for both $h_3$ and $h_4$. There is a ring-like region of $\approx$\,0 values at $\approx$\,20\,pc from the nucleus in the \textit{h$_{3}$} map indicating that the LOSVD in this region has a Gaussian profile. \textit{h$_{3}$} rises to $\approx$ 0.05 within a region with radius $\approx$\,17\,pc centered on the nucleus and also beyond the ring, indicating a skewness of the LOSVD towards the red in these regions. Low values of \textit{h$_{4}$} occur mainly in the nucleus (\textit{h$_{4}$}\,$\approx$\,0) and the highest along the minor axis (\textit{h$_{4}$}\,$\approx$\,0.06), indicating a LOSVD slightly more ``pointy'' than a Gaussian at the later locations.

We have modelled the centroid velocity field using circular orbits in a Plummer potential \citep{plummer11} as we have successfully done in previous works \citep{fausto06,riffel08}. This model is just a kinematic model and does not take into account the full LOSVD of the stellar components. The stars are assumed to orbit in circular orbits in a thin disk coplanar with the main galactic disk, in spite of the fact that the stellar velocity dispersion is not negligible. In addition we did not take into account the seeing and instrumental effects as our goal is only to derive the mean velocity field to be subtracted from the observed velocity field of the gaseous component.

The fit of the model gives a heliocentric systemic velocity of $-125\,\pm$11\,km\,s$^{-1}$ taking into account measurement errors in the fit, a kinematic center at horizontal and vertical coordinates of X\,=\,0.2\arcsec and Y\,=\,0.4\arcsec, respectively, relative to the peak flux in  the continuum (adopted as the center of the galaxy), and a position angle for the line of nodes of PA\,=\,153\ensuremath{^\circ}$\pm$1\ensuremath{^\circ}. The displacement between the kinematic and photometric  center is smaller than the seeing radius and it can thus be concluded that the two centers agree within the uncertainties. We have adopted a disk inclination of i\,=\,58\ensuremath{^\circ}, obtained from the aparent axial ratio (from NED\footnote{NASA/IPAC extragalactic database}) under the assumption of a thin disk geometry, because the inclination is not well constrained in our multi-parameter fitting procedure. The stellar centroid velocity field, the fitted circular model and the residuals between the stellar and modelled field are shown in the top panels of Fig.\ref{fig3}. The residuals are, at most locations, smaller than 10\,km\,s$^{-1}$, for a total velocity amplitude of $\approx$\,100\,km\,s$^{-1}$, revealing a good fit. Nevertheless, it can be noticed that the isovelocity curves of the observed velocity field show a small ``s-shape'' distortion when compared to the model, associated to the bluest and reddest residuals. In the bottom of Fig.\,\ref{fig3} we show the observed and model velocities extracted within a pseudo-slit $\approx$\,0.2\arcsec wide, oriented along the major axis at PA\,=\,153\ensuremath{^\circ}, in order to illustrate better the fit. The systemic velocity obtained from the fit V$_{s}$\,=\,$-$125\,$\pm$11\,km\,s$^{-1}$ was adopted as the galaxy systemic velocity and subtracted from the stellar and gaseous centroid velocities.

\subsection{Gaseous Kinematics}

\begin{figure*}
\includegraphics[scale=0.95]{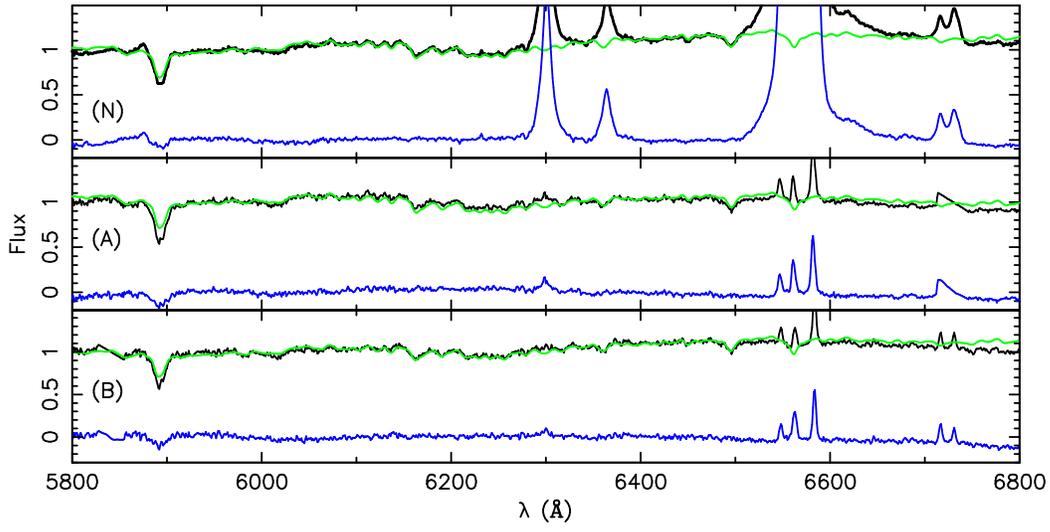}
\caption[Stellar Kinematics]{Observed spectrum (black), fitted absorption spectrum (green) and residual spectrum (blue) for three different regions.}
\label{fig4}
\end{figure*}

\begin{figure*}
\includegraphics[scale=1]{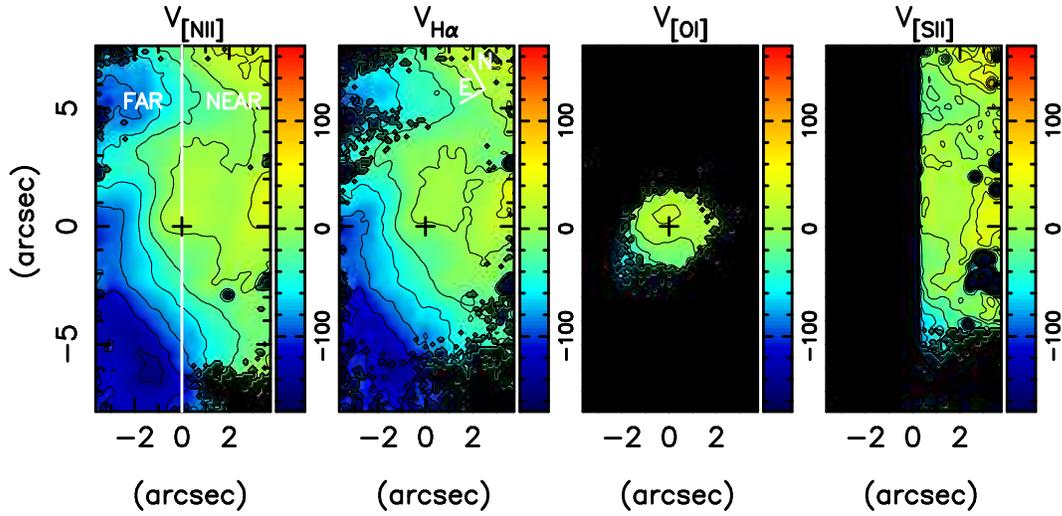}
\caption[Gaseous Kinematics]{Gaseous centroid velocity (km\,s$^{-1}$) for the [N\,II], H$\alpha$, [O\,I] and [S\,II] emission lines.}
\label{fig5}
\end{figure*}

\begin{figure*}
\includegraphics[scale=1]{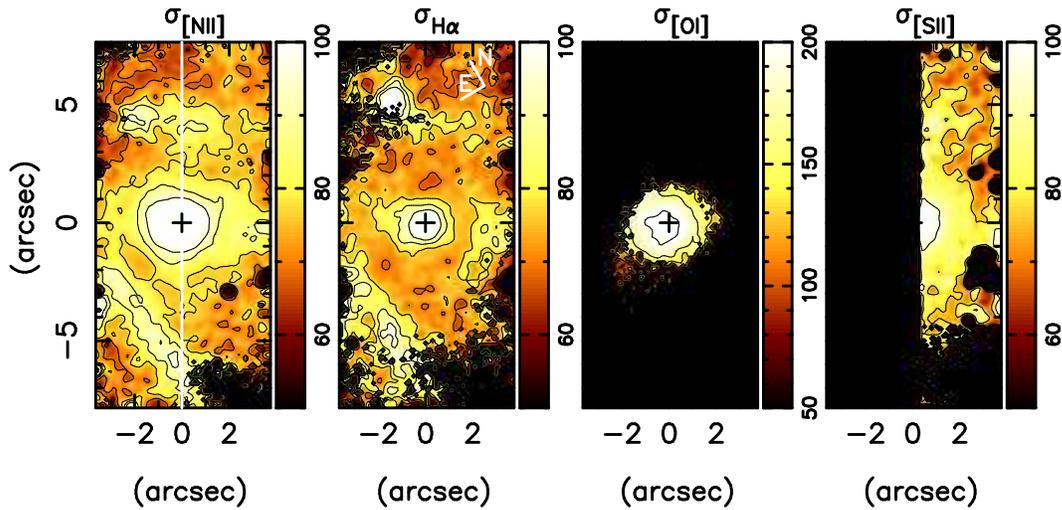}
\caption[Gaseous Kinematics]{Velocity dispersion (km\,s$^{-1}$) for the [N\,II], H$\alpha$, [O\,I] and [S\,II] emission lines.}
\label{fig6}
\end{figure*}

\begin{figure*}
\includegraphics[scale=1]{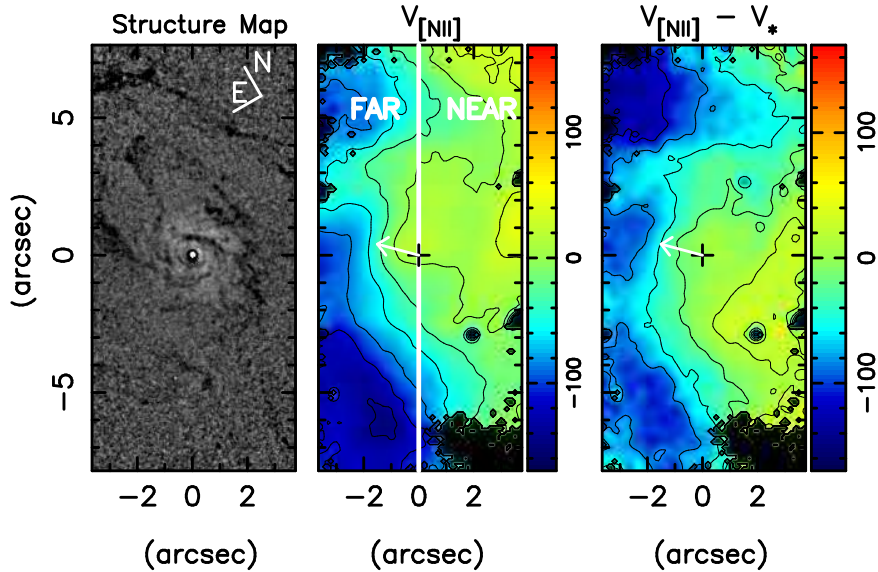}
\caption[Gaseous Kinematics]{From left to right: structure map, gaseous centroid velocity (km\,s$^{-1}$) and the residual between gaseous centroid velocity and stellar velocity field (km\,s$^{-1}$). The vertical white line in the central panel indicates the position of the line of nodes, while the arrow in the right panel shows the orientation of the compact radio jet.}
\label{fig7}
\end{figure*}

\begin{figure}
\includegraphics[scale=0.56]{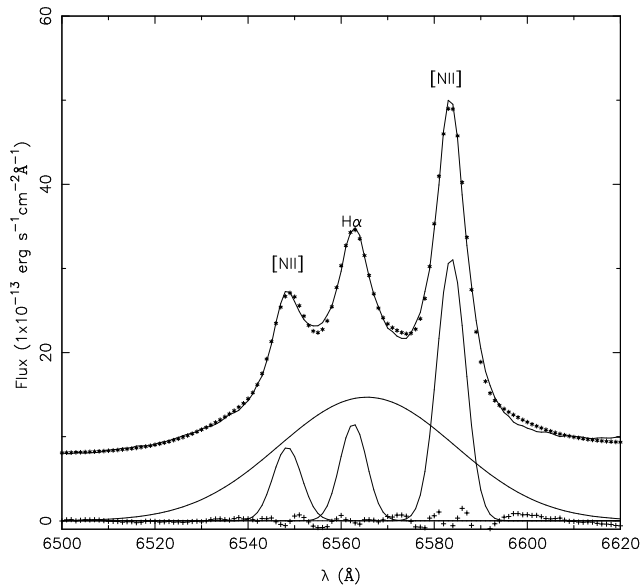}
\caption[Gaseous Kinematics]{Upper panel: Sample spectrum of the nucleus (black line) with the multi-Gaussian fit (represented by asterisks) to the emission lines superimposed. The individual Gaussian components are plotted below the spectrum along with the residuals between the spectrum of the nucleus and the fitted spectrum (crosses).}
\label{fig8}
\end{figure}

In order to measure the gas kinematics we have subtracted the contribution from the stellar population. This was done using the STARLIGHT software \citep{cid05}, in which the best-fitting absorption spectra is obtained through the technique of spectral synthesis. An illustration of the procedure is shown in Fig.\,\ref{fig4} for the nucleus and for locations A and B (Fig.\,\ref{fig1}).

The gas kinematics -- centroid velocities and velocity dispersions -- as well as the emission-line fluxes were obtained by the fit of Gaussians to the [N\,II], H$\alpha$, [O\,I] and [S\,II] emission lines in the subtracted spectra. We note that the [S\,II] line could only be measured in the half of the field where the line profile was not affected by the gap between the CCDs.

Errors were estimated from Monte Carlo simulations in which gaussian noise is added to the spectra. One hundred iterations were performed. Errors in the velocity, flux distribution and velocity dispersion measurements for the [N\,II] and H$\alpha$ emission lines have their lowest values within a radius of 2\arcsec\,\, and grow when approaching the borders of the field, where the signal-to-noise ratio is lower. The error values in both the velocity and velocity dispersion are respectively $\approx$\,10\,km\,s$^{-1}$ for the inner region and $\approx$\,20\,km\,s$^{-1}$ for the outer region. For the emission line fluxes they are respectively $\approx$\,10\% for the inner region and $\approx$\,25\% for the outer region. In the case of the [S\,II] emission lines, errors in the velocity and velocity dispersion are $\approx$\,15\,km\,s$^{-1}$ and in the fluxes $\approx$\,20\% over the whole FOV. For the [O\,I] emission line errors in the velocity and velocity dispersion are $\approx$\,10\,km\,s$^{-1}$ and $\approx$\,15\,km\,s$^{-1}$ respectively, while in the fluxes they are $\approx$\,15\%. Regions with errors in velocity higher than $\approx$\,20\,km\,s$^{-1}$ were masked out.

Within the inner 1.5\,\arcsec\,\,it was also necessary to fit a broad component to H$\alpha$. The need for this became obvious when we examined line ratio maps such as [N\,II]/H$\alpha$: within a circular region with radius of $\approx$\,1\,arcsec around the nucleus, the ratio decreased abruptely from values of $\approx$\,1.5 outside this region to values of $\approx\,$1 inside. By including a broad component within the inner region, the ratio between the narrow components [N\,II]/H$\alpha$ showed a smooth transition between the two regions. In addition, by including the broad component in the fit, the width of the narrow component of H$\alpha$ resulted practically the same as those of the [N\,II] emission lines, giving support to the presence of the broad component. As an additional  ``sanity check'' to the above procedure we also verified the resulting [N\,II]$\lambda\lambda$6583/6548 ratio which kept a value $\approx\,3.0\pm0.2$ over the whole region where the fit to the broad line was performed.

An ilustration of the fit of the [N\,II]\,$\lambda\lambda$6548,6583\,\r{A} lines and the broad and narrow H$\alpha$ components is presented in Fig.\,\ref{fig8} together with the residuals from the fit. Most of the flux of the broad component is well reproduced by a gaussian with $\sigma\,=\,862\pm37$\,km\,s$^{-1}$ and $V\,=\,-25\pm60$\,km\,s$^{-1}$

Centroid velocity maps for the emission lines are shown in Fig.\,\ref{fig5} and velocity dispersion maps are shown in Fig.\,\ref{fig6}. The velocity fields in the [N\,II] and H$\alpha$ are practically identical, and show that the gas is not in rotation along the same axis of the stars. If the gas were in rotation, the orientation would be perpendicular to that of the stellar rotation. Since the centroid velocities for the different emission lines are almost identical, we chose the [N\,II] centroid velocity to represent the gaseous velocity field due to its higher signal-to-noise ratio over most of the field. 

In order to isolate possible streaming motions we constructed a residual map by subtracting the stellar centroid velocity field from the gaseous one, under the assumption that the former represents the component of motion due to the galaxy gravitational potential. The residual map is shown in Fig.\,\ref{fig7} along with the structure map and the [N\,II] velocity field previous to the subtraction of the stellar velocity field. 

The H$\alpha$ and [N\,II] velocity dispersion maps (hereafter $\sigma_{H\alpha}$ and $\sigma_{[N\,II]}$) are very similar. In most of the field both $\sigma_{[N\,II]}$ and $\sigma_{H\alpha}$ present the same values (within errors), usually ranging from 60\,km\,s$^{-1}$ to 80\,km\,s$^{-1}$. An exception occurs inside a radius of $\approx$\,0.8\arcsec\,\, from the nucleus, where $\sigma_{[N\,II]}$ is about 20\,km\,s$^{-1}$ larger than $\sigma_{H\alpha}$ (uncertainties being $\approx$\,5\,km\,s$^{-1}$ in this region). Also inside this radius $\sigma_{[N\,II]}$ and $\sigma_{H\alpha}$ reach their highest values, \,$\approx$\,140\,km\,s$^{-1}$ and \,$\approx$\,130\,km\,s$^{-1}$ respectively. Away from the nucleus, an increase in the velocity dispersion is seen at two locations: $\approx$\,5\arcsec\,\,NW and $\approx$\,4\,\arcsec\,\,E from the nucleus, extending approximately diagonally in the IFU towards the S. These locations correspond to the largest blueshifts in the gas velocity fields.

\section{Line fluxes and excitation of the emitting gas}\label{fluxgas}

\begin{figure*}
\includegraphics[scale=1]{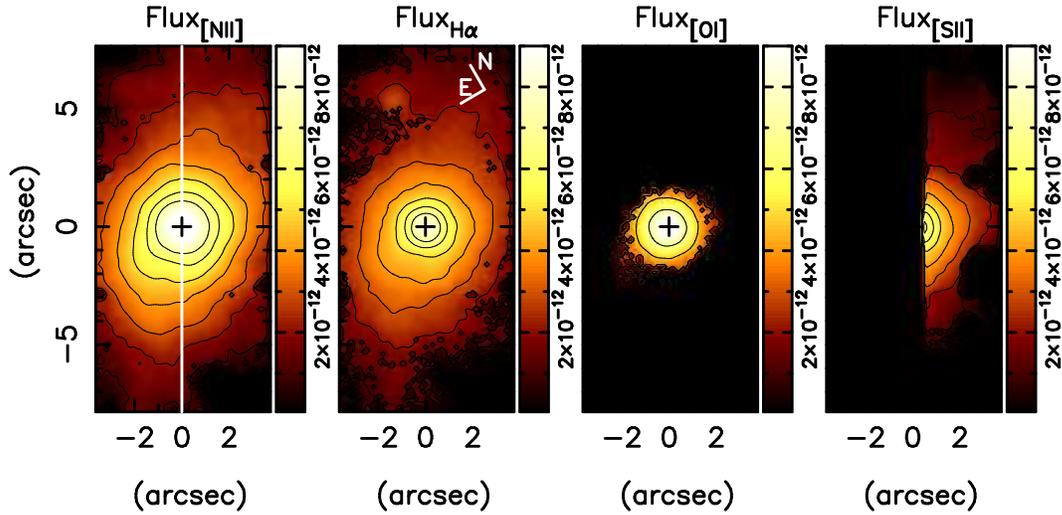}
\caption[Flux maps]{Maps of the [N\,II], H$\alpha$, [O\,I] and [S\,II] integrated fluxes (erg\,cm$^2$\,s$^{-1}$ per pixel) }
\label{fig9}
\end{figure*}

\begin{figure*}
\includegraphics[scale=1]{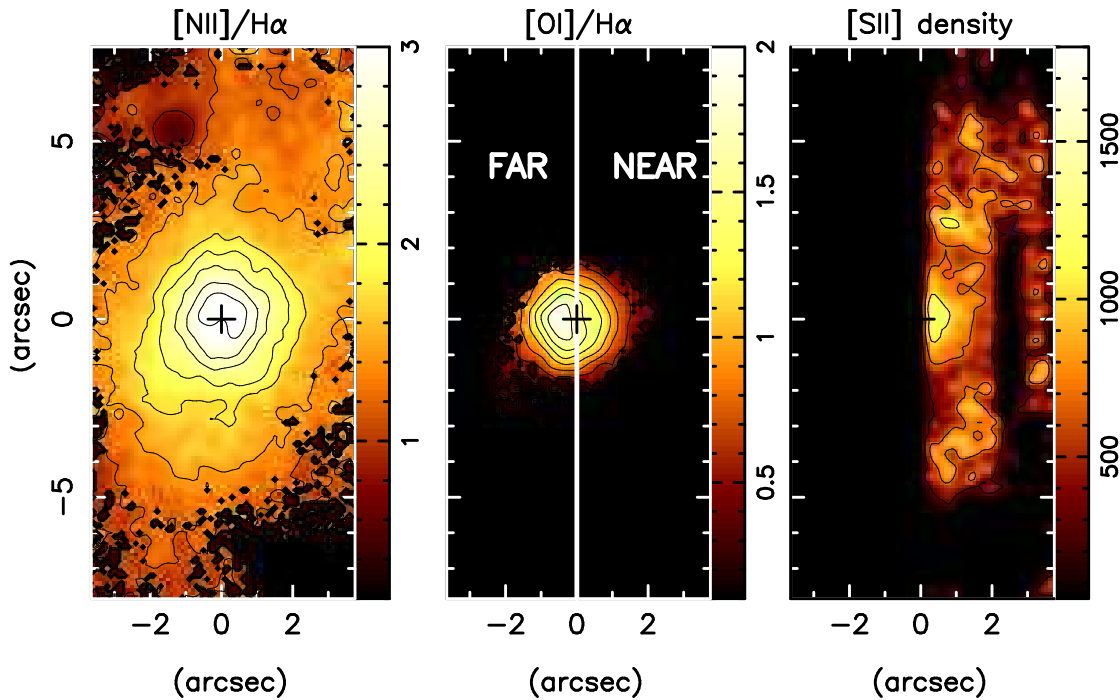}
\caption[Line ratio maps]{From left to right: [N\,II]/H$\alpha$ and [O\,I]/H$\alpha$ emission line ratios and the gas density (cm$^{-3}$) derived from the [S\,II] line ratio. }
\label{fig10}
\end{figure*}

In Fig.\,\ref{fig9} we present the integrated flux distributions for the [N\,II]\,$\lambda$6583\,\r{A}, the narrow component of H$\alpha$, [O\,I]\,$\lambda$6300\,\r{A} and [S\,II]\,$\lambda$6716\,\r{A} emission lines. While the [O\,I] flux distribution is compact, restricted to the inner $\approx$\,2\,arcsec around the nucleus, the remaining flux distributions show an elongation from NW to SE, with the H\,$\alpha$ flux distribution suggesting the presence of spiral arms at the top and bottom borders of the field. In the top left of the H$\alpha$ panel, to the north of the nucleus, there is a small region with enhanced emission, suggesting the presence of an HII region at this location.

In Fig.\,\ref{fig10} we show the line ratio maps [N\,II]/H$\alpha$ and [O\,I]/H$\alpha$ obtained using the narrow component of H\,$\alpha$, as well a gas density map, obtained from the [SII]\,$\lambda\lambda$6717/6731 line ratio assuming an electronic temperature of 10000K \citep{osterbrock89}. The [N\,II]/H$\alpha$ ratio is ``LINER-like'' -- with [N\,II] stronger than H$\alpha$ -- over most of the IFU field. The only region with smaller ratios, typical of HII regions is observed at the location of enhanced H$\alpha$ emission to the north of the nucleus, supporting the interpretation that there is an HII region there. The highest values are observed at the nucleus (where the [N\,II]/H$\alpha$ ratio is $\approx$\,3) and decrease outwards.

The [O\,I]/H$\alpha$ ratio (central panel of Fig.\,\ref{fig10}) presents values typical of AGN, but there is a gradient in the values which range from $\approx$\,1.0  at $\approx$\,1\arcsec\,\,SW of the nucleus to $\approx$\,1.9 at $\approx$\,0.5\arcsec\ to the NE of the nucleus along PA\,$\approx$\,50-60 (along the horizontal direction to the left in the figure). The asymmetry in the [O\,I]/H$\alpha$ ratio in the nuclear region seems to be mostly due to an asymmetry in the [O\,I] flux distribution.
 
The [S\,II] density reaches its highest value within the inner arcsecond (17\,pc), where it exceeds\,10$^{3}$\,cm$^{-3}$, decreasing to $\approx$\,500\,cm$^{-3}$ at 2 arcseconds (34\,pc) from the nucleus, the lowest measurable values are $\approx$\,300\,cm$^{-3}$ observed farther away.

\section{discussion}\label{discuss}

\subsection{Stellar Kinematics}

A previous study of the stellar kinematics  along PA\,=\,157\ensuremath{^\circ} by \citet{beltran01}, although on a much larger scale, has shown a smooth rotation pattern for the stars, which looks similar to ours, considering this specific PA. Nevertheless, our two-dimensional coverage allowed us to verify the presence of an s-shape distortion in the iso-velocity contours. \citet{elmegreen95} have  argued for the presence of a nuclear bar extending up to $\approx$\,0.5\,kpc (15\,arcsec) from the nucleus along PA\,$\approx$\,140$^\circ$, inferred from the fit of elliptical isophotes to a continuum J-band image of M\,81. This bar is thus oriented only 10$^\circ$  from the vertical direction in Fig.\,\ref{fig3} and extends beyond the FOV of our observations with reaches only 7$\farcs$5 along this PA. We may thus be sampling a region inside the bar and both the presence of an s-shape distortion in the iso-velocity contours and the x-shaped residuals in Fig.\,\ref{fig3}  are consistent with the elongated pattern of stellar orbits required to support a bar.

Considering an inclination of 58$^\circ$, the amplitude of the rotation curve to each side of the nucleus reaches at most 60\,km\,s$^{-1}$. This is much smaller than the velocity dispersion observed over the whole IFU field, which ranges from 140 to 190\,km\,s$^{-1}$. We thus can use the stellar velocity dispersion map (Fig. \,\ref{fig3}) alone to obtain a virial estimate for the mass of the bulge. The average value of the velocity dispersion inside a radius of 3.5\arcsec\ (the maximum covered by our observations along the shorter axis of the IFU) is 167\,km\,s$^{-1}$. In order to obtain the velocity dispersion of the bulge this value needs to be corrected according to equation (2) of \citet{jorgensen95}. Adopting an effective radius of r$_{e}$\,=\,66.2\arcsec\ \citep{baggett98} we obtain 162\,km\,s$^{-1}$, which is in good agreement with previous measurements \citep{beifiori09,beltran01,nelson95}. This leads to a virial mass of 2.1\,$\times$10$^{8}$\,M$_{\odot}$. 

The above bulge velocity dispersion can also be used to estimate the central black hole mass using the M-$\sigma$ relation \citep{tremaine02,gultekin09}, which results M$_{BH}$\,=\,5.5$_{-2.0}^{+3.6}$\,$\times$10$^{7}$\,M$_{\odot}$. This value is consistent with the previous virial estimate of the total mass, and is of the same order of the values obtained by \citet{devereux03} and \citet{beifiori09} for the mass of the black hole, which are 6.7$\,\times\,$\,10$^{7}$\,M$_{\odot}$ and 3.7$\,\times\,$\,10$^{7}$\,M$_{\odot}$ respectively when rescaled to the distance we adopted in this study.

\subsection{Gaseous Kinematics and Excitation}
\label{gaskinexc}

On kiloparsec scales, the H\,I gas kinematics  \citep{rots75} shows a similar rotation pattern to that of our stellar velocity field, and our  PA\,=\,$153\ensuremath{^\circ}$ of the line of nodes is comparable to the value of PA\,=\,150\ensuremath{^\circ} obtained by \citet{goad76}. Nevertheless, on the  scales probed by our observations, the gas kinematics is very different from that of the stars. This was already noticed by \citet{goad76}, who obtained the kinematics for the [N\,II]\,$\lambda$6583\,\r{A} and H$\alpha$ emission lines in the inner 1\,kpc of the galaxy and found the gaseous velocity field to be complex and impossible to describe by simple rotation alone. A disturbed kinematics was also found by \citet{beltran01} in the [O\,III]\,$\lambda$5007 emission-line gas. But they argue that their long-slit data along PA$=157^\circ$ suggest that there is a circumnuclear Keplerian disk of ionized gas. Our H\,$\alpha$ kinematics along this axis does not show this behaviour. Probably because [O\,III] and H\,$\alpha$ are probing distinct kinematics.

The gas isovelocity curves in our Fig.\,\ref{fig5}  suggest a kinematic major axis almost perpendicular to that of the stars and even more than one rotation center. Nevertheless, the location of these apparent rotation centers, as well as the isovelocity contours in their vicinity, correlate with the locations of increased velocity dispersions (Fig.\,\ref{fig6}) suggesting that these regions are related to shocks in the gas. The overall orientation of these features seem also to be the same as those of the dust lanes in the structure map. We also note that most blueshifts are in the far side of the galaxy, suggesting inflow towards the center if most of the gas is in the galaxy plane. In this case the shocks mentioned above could be responsible for dissipating angular momentum and allowing the gas to flow in towards the nucleus to feed the AGN. If a nuclear bar is present, as suggested by the photometry of \citet{elmegreen95} and by the distortions of the stellar isovelocity curves, this bar could also play a role in channeling gas inwards from the outer regions of the galaxy.  This possibility is supported by the gas flux maps (Fig.\,\ref{fig9}), which are elongated towards the bar orientation.

If the gas is not in the galaxy plane, the observed blueshifts could be due to outflow to high galactic latitudes. Relevant to this hypothesis is the observation of a compact one-sided radio jet by \citet{bietenholz00} and, more recently, by \citet{markoff08}. This jet is seen at an average position angle of PA\,=\,50\ensuremath{^\circ}\,$\pm$\,6\ensuremath{^\circ} and extends only to $\approx\,0\farcs$003 from the nucleus \citep{markoff08}. We plotted its direction on Fig\,\ref{fig7}. Although it points to some blueshifts observed beyond $\approx$\,1$\farcs$5 to the left of the nucleus in the Figure, the gas closer to the nucleus (thus at the scale of the jet extent) is not observed in blueshift, suggesting that the blueshifts are not related to the radio jet.

Within the inner arcsecond around  the nucleus, the increase in the gas velocity dispersion  may be related to the compact nuclear jet at PA\,=\,50$^\circ$ in interaction with the surrounding gas. Coincidently, this is approximately the PA along which we observe an increase in the [OI]/H$\alpha$ line ratio (Fig.\,\ref{fig10}), a well known signature of shocks \citep{bpt81,osterbrock89}.

\subsection{Estimating the mass accretion flow}
\label{inflow}

The residual blueshifts observed in Fig.\,\ref{fig7} reach high values of $\approx$\,$-$100\,km\,s$^{-1}$ in three regions in the far side of the galaxy. If we assume that the gas is located in the galaxy plane, this means that the gas is inflowing to the nucleus and we can thus estimate the rate of mass inflow. In order to do this, we assume that the three regions channel gas towards the nucleus. Assuming a similar geometry for the three regions, the total gas mass inflow will be three times the flow in one region. The flux of matter crossing one region can be calculated as:
\[
\phi\,=\,N_{e}\,v\,\pi\,r^{2}\,m_{p}\,f
\]
where $N_{e}$ is the electron density, $v$ is the streaming velocity of the gas towards the nucleus, $m_{p}$ is the mass of the proton, $r$ is the cross section radius of the inflow and $f$ is the filling factor. For the purpose of the present calculation, we adopt a cross section radius of 17\,pc and a typical value for the filling factor of 
$f$\,=\,0.001.

The velocity residuals in the three blueshifted regions on the far side of the galaxy, which we have interpreted as due to streaming motions, range from $\approx$\,$-80$\,km\,s$^{-1}$ to $\approx$\,$-120$\,km\,s$^{-1}$ (Fig.\,\ref{fig7}). As we have assumed that these streaming motions occur in the plane of the galaxy, we need to correct these values for the inclination of the galaxy i\,=\,58.4\ensuremath{^\circ}. The resulting average velocity for the streaming motions is  $v$\,=\,$-$117\,km\,s$^{-1}$. For $N_{e}$\,=\,500\,cm$^{-3}$ (obtained from the [S\,II] ratio) and $v$\,=\,$-$117\,km\,s$^{-1}$, we obtain the value for the total inflow of ionized gas mass of $\phi$\,$\approx$\,4.0\,$\times$\,10$^{-3}$\,M$_{\odot}$\,yr$^{-1}$.

We note that the above calculation is only a rough estimate, as it depends on uncertain parameters, such as the geometry of the flow. We point out also the fact that the estimated inflow rate is in ionized gas and is thus a lower limit to the total inflow rate, as it may be only the ionized ``skin'' of a  more massive inflow in non-emitting neutral gas. The presence of molecular gas closer than $\approx$\,300\,pc from the nucleus seems to be ruled out by the observations of \citet{sakamoto01}, who found that molecular gas is mainly on a ``pseudoring'' or spiral arm at about 500\,pc from the nucleus.

We now compare the estimated inflow rate of ionized gas to the mass accretion rate necessary to produce the luminosity of the LINER nucleus of M\,81, calculated as follows:
\[
\dot{m}\,=\,\frac{L_{bol}}{c^{2}\eta} 
\]
where $\eta$ is the efficiency of conversion of the rest mass energy of the accreted material into radiation. For LINERs it has been concluded that, in most cases, the accretion disk is geometrically thick, and optically thin \citep{nemmen06, yuan07}. This type of accretion flow is known as RIAF (Radiatively Inefficient Accretion Flow \citep{narayan05}), and has a typical value for $\eta$\,$\approx$\,$0.01$. The nuclear luminosity can be estimated from the X-ray luminosity of $L_{X}$\,=\,3.2\,$\times\,$10$^{40}$\,erg\,s$^{-1}$ \citep{ho99}, using the approximation that the bolometric luminosity is $L_{B}$\,$\approx$\,10$L_{X}$. We use these values to derive an accretion rate of $\dot{m}$\,=\,5.65$\,\times\,$10$^{-4}$\,M$_{\odot}$\,yr$^{-1}$.

Comparing this accretion rate $\dot{m}$ with the mass flow of ionized gas along the the three regions derived above $\phi$, we conclude that $\phi$\,$\approx$\,7\,$\dot{m}$. Thus the inflow of ionized gas in the inner tens of parsecs of M\,81 is already of the order of the one necessary to feed its AGN.

\subsection{PCA Tomography}\label{pca}

\begin{figure*}
\includegraphics[scale=1]{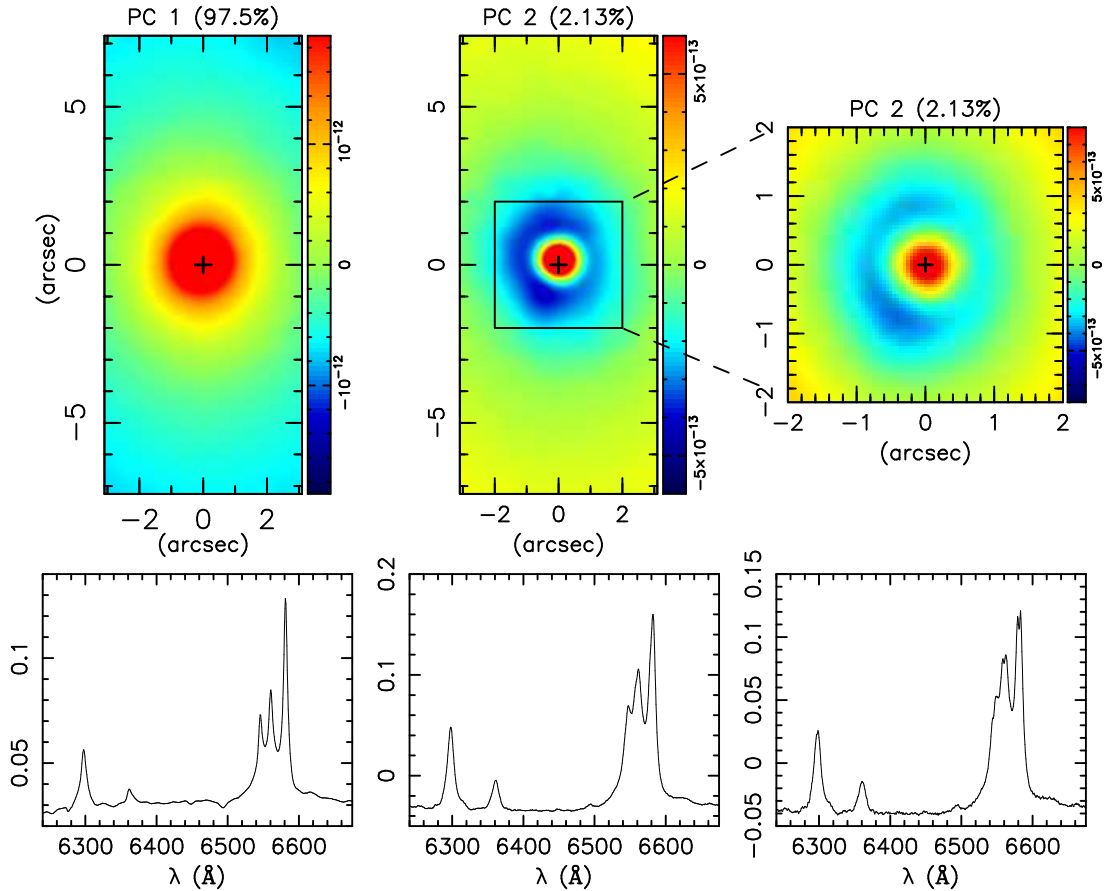}
 \caption{Left: first eigenspectra (bottom) ad its respective tomograms (top); center and right: second eigenspectra and tomogram resultant from the application of PCA to the larger (central panels) and smaller (right panels) spatial regions. The square in the central panel shows the smaller central region where the PCA was also applied.}
\label{fig11}
\end{figure*}

\begin{figure*}
\includegraphics[scale=1]{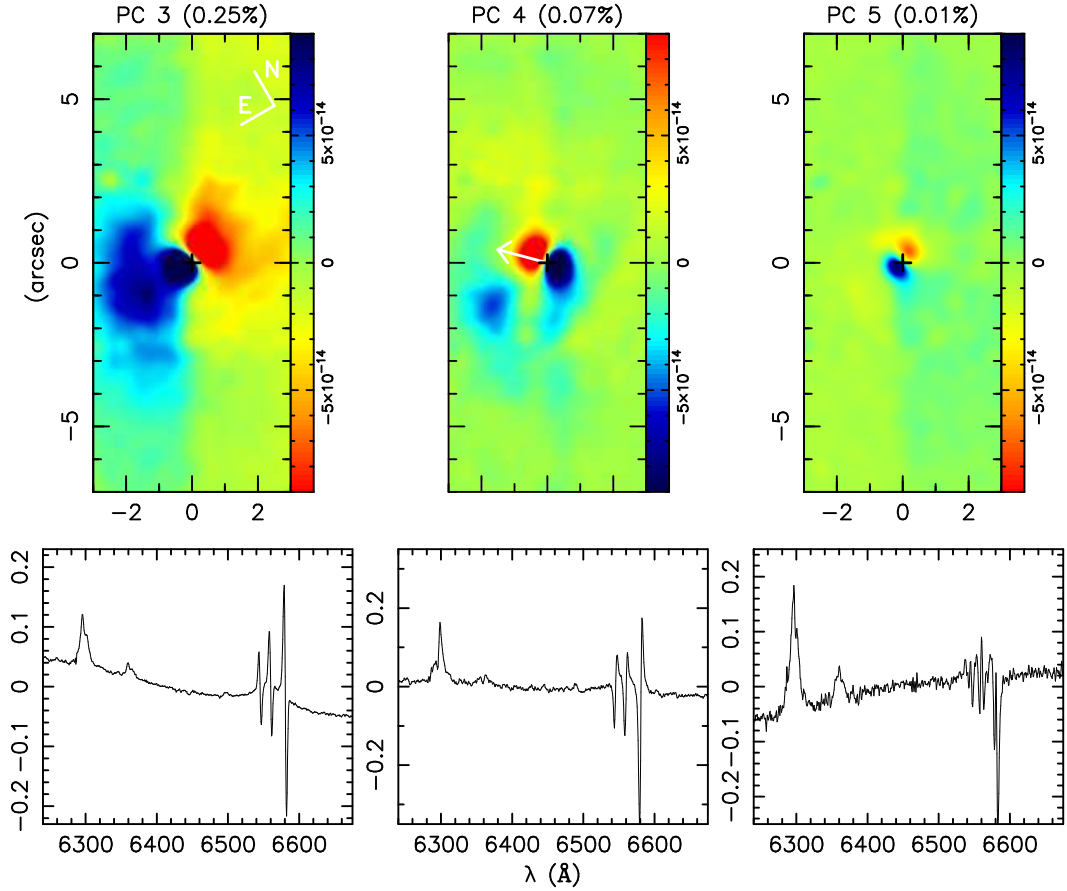}
\caption{Bottom: third, fourth and fifth eigen-spectra; top: corresponding tomograms. The correlations/anticorrelations in both eingenvectors and tomograms show gas movement and the approaching/receding sides can easily be identified.}
\label{fig12}
\end{figure*}

\begin{figure*}
\includegraphics[scale=1]{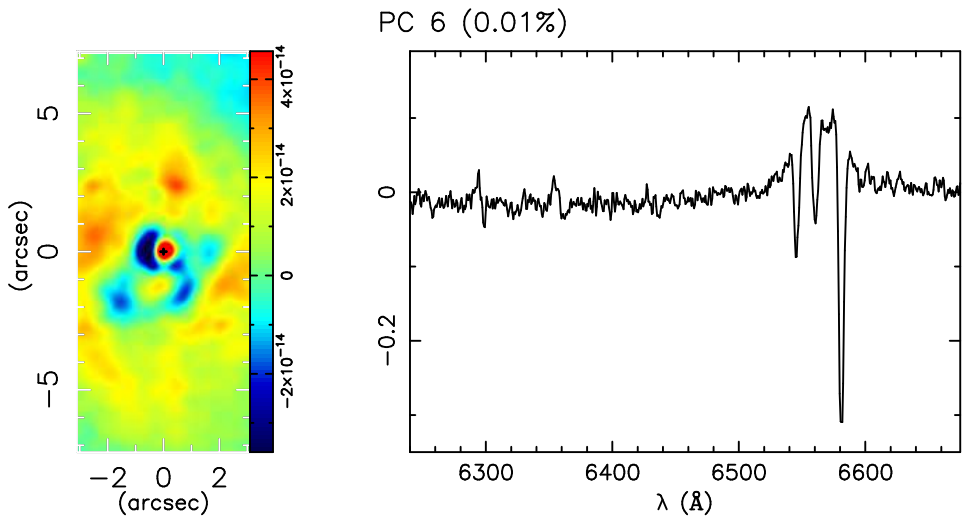}
\caption{Left: Tomogram of PC\,6; right: eigenvector 6, showing the broad H$\alpha$ component.}
\label{fig13}
\end{figure*}

In order to try to extract more information from the complex gas kinematics of the nuclear region, we have used an alternative approach which has been successfully applied to reveal small scale features around nearby AGN by \citet{steiner09}: the method of Principal Component Analysis. The method, which is applied directly to the calibrated data cube, allows the separation of the information, originally presented in a system of correlated coordinates, into a system of uncorrelated coordinates ordered by principal components of decreasing variance. These new coordinates called ``eigenspectra'' (eigenvectors which are functions of wavelength) reveal spatial correlations and anti-correlations associated to features that may be associated to emission-line regions. The projection of the data onto these new coordinates produce images called ``tomograms'', which represent slices of the data in the space of the eigenvectors. In order to apply this technique, we first corrected the data for differential atmospheric refraction and applied a Richardson-Lucy deconvolution algorithm (Richardson 1972; Lucy 1974) to the data cube using 6 iterations. A larger number of iterations did not improve the results, as also verified in previous works. A discussion about this point can be found in \citet{morelli2010} and \citet{michard96}. The real PSF shows a small variation with wavelenght and a Gaussian with FWHM\,=\,0.6\,\arcsec\,\,was found to be a good representation of the average PSF over the different wavelenghts. We then applied the PCA to the inner 6\,$\times$\,14\,arcsec$^{2}$ and, in order to enhance the features in the vicinity of the active nucleus, also to the inner 4\,$\times$\,4\,arcsec$^{2}$ of the galaxy. We restricted the spectral range to $\lambda\lambda$\,6200-6700\,\r{A}, as there is no relevant information on the gas emission for $\lambda\,<\,6200$\AA\, and the sulphur lines at $\lambda\lambda$\,6717,6731 are affected by falling in the gaps of the CCD, as previously explained.

The resultant eigenspectra and respective tomograms did not change significantly when the analysis was performed on either the inner 6\,$\times$\,14\,arcsec$^{2}$ or 4\,$\times$\,4\,arcsec$^{2}$ of the galaxy, the second eigenspectra and tomogram being an exception, as they presented small but significant changes. Because of the similarity we choose to primarily focus on the results corresponding to the larger spatial region for all eigenspectra and tomograms except for the second, for which both are shown. 

We have performed a ``scree test'' (see \citet{steiner09}) to select the significant components -- the ones not dominated by noise, which we concluded are the first 8 (PC\,1 to PC\,8). The first six eigen-spectra and corresponding tomograms are shown in Figs.\,\ref{fig11}, \ref{fig12} and \ref{fig13}. PC\,7 and PC\,8, although significant, do not show any particular new structure.

Most of the variance is in the first eigenspectrum PC\,1 (principal component 1, which contributes with 97.50\% of the variance), which is similar to the spectrum of the nucleus plus central stellar bulge (bottom left panel of Fig.\ref{fig11}). The second tomogram, PC\,2 (with 2.13\% of the variance), shows an anticorrelation between the nucleus and the stellar bulge. In the eigenspectrum 2 (bottom central panel of Fig.\ref{fig11}) one can clearly see a pure nuclear emission spectrum (positive values) anticorrelated with the stellar emission (negative values) -- also evidenced by the 6500\,\r{A} stellar absorption feature, which appears in absorption in eingenspectrum 1 but appears inverted in the eigenspectrum 2. A somewhat conspicuous red H$\alpha$ wing may be reminiscent of the double-peaked broad H$\alpha$ reported by \citet{bower96}. This feature is certainly weaker than reported by \citet{bower96} and the stellar feature at 6500\,\r{A} should not be confused with emission from the blue H$\alpha$ wing. When the analysis is performed on the inner 4\,$\times$\,4\,arcsec$^{2}$ (right panels of Fig.\,\ref{fig11}) a weak feature not seen in the centroid velocity maps appears: double peaks in the emission lines of H$\alpha$+[N\,II], with the blue and red peaks slightly blueshifted ($\approx$\,$-$80\,km\,s$^{-1}$) and slightly redshifted ($\approx$\,80\,km\,s$^{-1}$), respectively, relative to the systemic velocity. The fact that the two peaks are correlated means that they are spatially coincident, that is, they come from the same region and are not spatially resolved. This can be interpreted as due to a compact bipolar outflow or rotation, not resolved in the data.

\subsubsection{Gas rotation versus outflow: feeding \& feedback} 

The third, fourth and fifth eigenspectra, PC\,3 (with 0.25\% of the variance), PC\,4 (0.07\%) and PC\,5 (0.01\%), show peaks associated to the H$\alpha$+[N\,II] emission lines spatially anti-correlated with symmetric lobes (Fig.\,\ref{fig12}). The pattern in the tomogram of PC\,5 is, in fact, very similar to that of PC\,3, and we thus consider that PC\,3+PC\,5 correspond to one structure and PC\,4, whose geometry in the tomogram is perpendicular to that of PC\,3+PC\,5, to another structure. These patterns can be explained either by a rotating gaseous disk or, else, by outflow of ionized gas, either a jet or a milder accretion disk outflow. A gaseous disk with such a behavior was already proposed for NGC\,4637 by \citet{steiner09}. In the present case, however, we see two such structures with axes approximately ortogonal to each other. This can be explained by a simple model of a rotating disk and an outflow driven by this disk or by an unresolved accretion disk inside. This explains not only the existence of the two patterns, seen in PC\,3+PC\,5 and PC\,4, but also the reason why they are orthogonal.

The immediate question that arises is which configuration, PC\,3+PC\,5 or PC\,4, is associated to the rotating gaseous disk and which one is produced by the outflow. We propose that PC\,3 and PC\,5 are associated to the gaseous disk and PC\,4 to the outflow. This interpretation is corroborated by the overall gas rotation seen in the galaxy as shown by the [N\,II] velocity field displayed in Fig.\,\ref{fig7}. The rotation axis seen in this figure (close to the nucleus) approximately coincides with the one seen in Fig.\,\ref{fig12} for PC\,3+PC\,5 as does the sense of redshift-blueshift rotation. The PC\,3 tomogram shows that the disk is within $\approx$\,50\,pc from the nucleus and comparison of this tomogram with the gas kinematics of Fig.\,\ref{fig7} suggests that the disk is being fed  by the gas circulating in the inner part of the galaxy. A compact disk in rotation around the nucleus has also been found in the H$_2$ kinematics of the inner few tens of parsecs of NGC\,4151 \citep{thaisa10} and Mrk\,1066 \citep{rogemar11}.

Why do we see two eingenvectors, PC\,3 and PC\,5, with similar spectral structures? They are not really similar. In PC\,3 the [O\,I] emission is seen in correlation with a bluish continuum while in PC\,5 this emission is correlated with a reddish continuum. They very probably are associated to two regions affected  by different dust reddening or, else, by distinct scattering properties, or both. Thompson scattering by free electrons is wavelength independent but dust scattering or molecular Rayleigh  scattering enhances blue characteristics. The partially ionized region responsible for the [O\,I] emission seen in tomogram 3 could be exposed to the central blue continuum source and scatter it to our line of sight, enhancing the blue aspect of the continuum emission.

We interpret PC\,4 as due to an outflow, which most probably originates in the interaction of the radio jet discussed in Sec.\,\ref{gaskinexc} with the circumnuclear gas. We plotted the direction of the radio jet on tomogram 4 (arrow in Fig\,\ref{fig12}). Considering that \citet{markoff08} argues that this jet precesses within 20$\degr$ from this orientation, the jet direction approximately coincides with  that of the axis of the presumed disk in tomogram 3. This is what one would expect if a conical outflow or jet is induced by an inner accretion disk; assuming that the axis of this inner accretion disk is aligned with the structure seen in tomogram 4, we have a self-consistent picture. As the highest energy photons are probably escaping perpendicular to the disk, we further predict that high ionization species, not present in the spectral range of our observations, such as [O\,III], [Ne\,III] or He\,II may be present in the blobs associated to PC\,4.

Tomograms 3 and 4 also show that the [O\,I] emission is spatially correlated with the gas emission to the E-N of the nucleus. This is revealed in the eigenspectrum 3 by the fact that [O\,I] is correlated with the blueshifted emission (to the east), while in the eigenspectrum 4 it is correlated with the redshifted emission (to the north east), in agreement with the previous result that the [O\,I] emission is enhanced towards the radio jet orientation of PA\,$\approx$\,50\ensuremath{^\circ}.  In addition, the [O\,I] profile has indications of double structure, being slightly redshifted. Although this question needs further studies we wonder whether this asymmetric [O\,I] enhancement is not related to the one-sided radio jet, perhaps due to shock heating of the partially ionized gas.

Finally, we show in Fig.\,\ref{fig13} the tomogram (left panel) and eingenspectrum (right panel) of PC\,6. The eigenspectrum shows that the unresolved broad component of H$\alpha$ (see Fig.\,\ref{fig8}), which is confined to the red circle at the nucleus in the tomogram, is spatially anti-correlated with the narrow-emission lines, which originate in the blue patches surrounding the nucleus.

\section{conclusions}\label{conc}

We have measured the kinematics of the stars and ionised gas in the inner 120\,$\times$\,250\,pc$^2$ of M\,81 using integral field spectroscopy obtained with the GMOS instrument at the Gemini North telescope with a spatial resolution of 10\,pc at the galaxy. The main results of this paper are:
\begin{itemize}

\item The stellar velocity field (centroid velocity) shows circular rotation with a major axis orientation of $\approx$\,153$^\circ$, which is similar to that of the large scale stellar rotation (as reported in previous works). But the two-dimensional coverage allowed the detection of deviations from pure rotation which can be  attributed to stellar motions possibly associated to a nuclear bar \citep{elmegreen95};

\item The stellar velocity dispersion of the bulge is 162\,$\pm$\,15\,km\,s$^{-1}$, in good agreement with previous measurements, resulting in an estimate of the black hole mass of M$_{BH}$\,=\,5.5$_{-2.0}^{+3.6}$\,$\times$10$^{7}$\,M$_{\odot}$;

\item The gaseous velocity field is completely distinct from the stellar one, and is dominated by blueshifts in the far side of the galaxy and a few redshifts in the near side. The subtraction of the stellar velocity field from the gaseous one confirms the presence of excess blueshifts of $\approx$\,$-$100\,km\,s$^{-1}$ on the far side and some redshifts in the near side. If the gas is in the galaxy plane, these excesses can be interpreted as streaming motions towards the center of the galaxy; 

\item From the measured gas velocities and assumed geometry, we estimate an ionized gas mass flow rate of $\phi$\,$\approx$\,4.0\,$\times$\,10$^{-3}$\,M$_{\odot}$\,year$^{-1}$, a value about seven times the necessary accretion rate to power the AGN. However, this is only the mass flow rate in ionized gas. Most probably, the total inflow is much larger but is dominated by non-emitting gas; 

\item We have applied the technique of principal component analysis (PCA) to our datacube and find that the PCA reveals small scale features not seen in the measured velocity field. The main features are: (1) an unresolved bipolar outflow or rotation at the nucleus; (2)  a nuclear rotating disk within $\approx$\,50\,pc from the nucleus; (3) a compact bipolar outflow approximately perpendicular to the disk;

\item The gas velocity field suggests that the nuclear disk is being fed by gas circulating in the galaxy plane;

\item An increase in the [O\,I]/H$\alpha$ line ratio occurs at approximately the same orientation of a one-sided nuclear radio jet (PA\,$\approx$50\ensuremath{^\circ}), which is also the approximate orientation of the bipolar outflow revealed by the PCA. The increase in the [O\,I]/H$\alpha$ line ratio is thus probably due to the interaction of the jet with the circumnuclear gas, as this ratio is a well known signature of shocks. 
\end{itemize}

With our observations we are thus resolving both the feeding of the M\,81 AGN -- via the compact disk and surrounding gas which seems to be replenishing the disk, and its feedback -- via the compact outflow approximately perpendicular to the disk, probably driven by this disk or by an unresolved accretion disk inside.

\section*{ACKNOWLEDGMENTS}

We acknowledge the referee for relevant suggestions which have improved the paper. This work is based on observations obtained at the Gemini Observatory, which is operated by the Association of Universities for Research in Astronomy, Inc., under a cooperative agreement with the NSF on behalf of the Gemini partnership: the National Science Foundation (United States), the Science and Technology Facilities Council (United Kingdom), the National Research Council (Canada), CONICYT (Chile), the Australian Research Council (Australia), Minist\'erio da Ci\^encia e Tecnologia (Brazil) and south-eastCYT (Argentina). This work has been partially supported by the Brazilian institution CNPq.

\bibliographystyle{mn2e.bst}
\bibliography{m81v10.bib}

\label{lastpage}
\end{document}